\documentclass[%
 aip,
 amsmath,amssymb,
 reprint,%
]{revtex4-1}
\usepackage{lineno}
\usepackage{graphicx}
\usepackage{dcolumn}
\usepackage{bm}
\usepackage[]{lineno}

\usepackage[utf8]{inputenc}
\usepackage[T1]{fontenc}
\usepackage{mathptmx}
\usepackage{etoolbox}
\usepackage{xcolor}
\usepackage{color,soul}
\makeatletter
\def\@email#1#2{%
 \endgroup
 \patchcmd{\titleblock@produce}
  {\frontmatter@RRAPformat}
  {\frontmatter@RRAPformat{\produce@RRAP{*#1\href{mailto:#2}{#2}}}\frontmatter@RRAPformat}
  {}{}
}%
\makeatother
\begin{document}
\preprint{AIP/123-QED}

\title[Wireless Josephson parametric amplifier above 20 GHz]{Wireless Josephson parametric amplifier above 20 GHz}
\author{Z. Hao}
\author{J. Cochran}
\author{Y.-C. Chang}
\author{H. M. Cole}
\author{S. Shankar}
\affiliation{Chandra Department of Electrical and Computer Engineering, University of Texas at Austin, 2501 Speedway, Austin, TX 78712, USA}
\email{shyam.shankar@utexas.edu}

\date{\today}

\begin{abstract}
Operating superconducting qubits at elevated temperatures offers increased cooling power and thus system scalability, but requires suppression of thermal photons to preserve coherence and readout fidelity. This motivates migration to higher operation frequencies, which demands high-frequency amplification with near-quantum-limited noise characteristics for qubit readout. Here, we report the design and experimental realization of a wireless Josephson parametric amplifier (WJPA) operating above 20~GHz. The wireless design eliminates losses and impedance mismatches that become problematic at high frequencies. The WJPA achieves more than 20~dB of gain across a tunable frequency range of 21--23.5~GHz, with a typical 3dB bandwidth of 3~MHz. Through Y-factor measurements and a qubit-based photon number calibration, we show that the amplifier exhibits an added noise of approximately two photons. 
\end{abstract}
\maketitle

Conventional superconducting qubit platforms for large-scale quantum computation~\cite{google2025,Gao2025establishing}  employ sub-10~GHz qubits operating below 20~mK in dilution refrigerators. At these temperatures, limited cooling power presents a significant challenge to scalability. Raising the operating temperature to the range of 100~mK–1~K offers substantially greater cooling headroom, enabling the operation of more qubits. However, this increase in temperature also elevates the excited-state population of sub-10~GHz qubits, necessitating more complex state preparation protocols such as autonomous cooling\cite{Magnard2018fast} or measurement-based reset\cite{Johnson2012herald}. Additionally, thermal population in sub-10~GHz readout resonators contributes to qubit dephasing via photon shot noise\cite{krantz2019quantum}.

To mitigate these effects, circuit QED systems operating at elevated temperatures must also transition to higher frequencies—typically in the 20–100~GHz range. Recent studies have demonstrated qubits operating above 20~GHz using advanced Josephson junctions and weak-links with superconducting Nb~\cite{anferov2024super,anferov2024millimeter} and TiN~\cite{purmessur2025operation}. In parallel, readout resonators are also being pushed to higher frequencies to suppress thermal photon-induced dephasing\mbox{\cite{anferov2024super}}. High-frequency resonators coupled to conventional frequency qubits are also under active investigation, as large detuning between qubit and readout frequencies can reduce measurement-induced state transitions, thereby enhancing readout fidelity and repeatability~\cite{Swiadek2024enhancing,Nesterov2024measurement,kurilovich2025high}. These developments collectively call for the development of ultra-low-noise amplification systems operating above 20~GHz to enable fast, high-fidelity qubit readout.

Kinetic-inductance traveling-wave parametric amplifiers (KITWPAs)  have demonstrated amplification up to 34~GHz for astronomy applications\mbox{\cite{shu2021nonlinearity,Tan2024operation}}. However, their added noise was not characterized and they are not in widespread use for superconducting qubit measurement.
Instead, state-of-the-art superconducting qubit measurement systems typically employ Josephson parametric amplifiers (JPAs) as the first stage of nearly quantum-limited amplification~\cite{aumentado2020,Roy2016}. However, conventional JPA designs operating below $\sim$12~GHz face significant challenges when scaled to higher frequencies. Standard packaging techniques—such as wirebonding JPA chips to printed circuit boards (PCBs)—introduce loss and impedance mismatches that worsen at high frequencies. Moreover, parasitic modes in these packages can interfere with the JPA mode and degrade noise performance. An example of such issues is mentioned in the high-frequency JPA work by Yurke \textit{et al.}\cite{Yurke1989} which presented a JPA operating at 23.5 GHz using Niobium-aluminum oxide Josephson junctions. While some of these limitations can be mitigated~\cite{anferov2024super}, scaling beyond 20~GHz is more readily achieved with designs that eliminate wirebonds and lossy PCB materials altogether. A promising precedent was established by Narla \textit{et al.}~\cite{narla2014}, who demonstrated a wireless JPA coupling scheme using a rectangular waveguide to realize parametric amplification at 10.2~GHz and below. In this work, we show that such wireless coupling schemes can be extended to realize parametric amplification at higher frequencies required for next-generation circuit QED systems.

 We report the design and characterization of a wireless Josephson parametric amplifier (WJPA) that operates in the K-band (18--26~GHz). The device, realized with aluminum/aluminum-oxide JJs, achieved more than 20~dB of gain over a tunable frequency range of 21 to 23.5~GHz, with an average 3dB bandwidth of 3~MHz. The WJPA, operated as a single pump four-wave mixing amplifier, achieved a 1~dB input compression power at 20~dB gain of $P_{\text{1dB}} = -125$~dBm. Using Y-factor and qubit-based photon number calibration techniques, the system added noise with the WJPA was measured to be approximately two photons at 22~GHz. 

\begin{figure}
    \centering
    \includegraphics[width=1\linewidth]{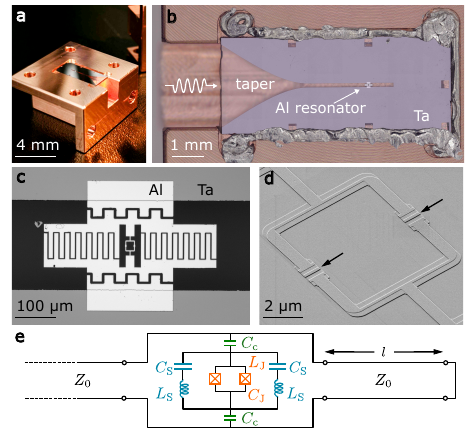}
    \caption{\textbf{High-frequency WJPA device and package.}
(a) Photograph of the copper waveguide package with the sapphire chip mounted in one half of the rectangular cavity. 
(b) Close-up view of the assembled chip inside the copper package.
(c) Optical image of the resonator region showing a lumped-element resonator consisting of an Al-AlO$_x$ JJ SQUID shunted by interdigitated finger capacitors, also coupled to the slotline ground planes (tantalum) via capacitors.  
(d) Electron micrograph showing the central SQUID loop. Arrows indicate the junctions. 
(e) Simplified circuit diagram of the WJPA. 
}
    \label{fig:device}
\end{figure}

The WJPA device is shown in Fig.~\ref{fig:device}(a-d), with equivalent circuit diagram depicted in Fig.~\ref{fig:device}(e). The device is realized by a planar on-chip circuit mounted in a 3D copper package, with microwave signals applied through a rectangular waveguide launch. The rectangular waveguide dimensions were chosen to be 10.67 mm by 4.32 mm, corresponding to the WR42 standard rectangular waveguide specification, for frequency range of 18 to 26~GHz (K-band). The package is designed to be attached to a commercial WR42-2.92mm coaxial cable adapter (not shown) which is then connected to other microwave components. Fig.~\ref{fig:device}(b) shows a top view of the chip sitting in the copper package where the left opening mates with the WR42 adapter. Inspired by previous work on millimeter-wave superconducting circuits~\cite{anferov2024low,anferov2024millimeter}, this design avoids wirebonds by using a lithographically defined transition from the waveguide to a slotline transmission line. The transition takes the form \( y(x) = \frac{W_a - S}{2} \cdot \frac{x}{A} \sqrt{2 - \left(\frac{x}{A}\right)^2} \) from Anferov \textit{et al.}~\cite{anferov2024low},
where \(W_a\) is the smaller rectangular waveguide dimension (4.32 mm for WR42), $S$ is the slotline gap width, and $A = 4.5$ mm is the transition length in our optimized design. An illustration of these geometric dimensions is depicted in the Supplementary Fig. S1(a). With electromagnetic simulation in Ansys HFSS, the taper was found to have a return loss below 10~dB and insertion loss under 1~dB over the target frequency band of 18 to 26 GHz. The slotline, defined on a sapphire substrate with superconducting ground planes, was designed with a gap width $S= 200$~$\mu$m, yielding a characteristic impedance of approximately $110~\Omega$. The slotline is shorted at one end, and a tunable lumped-element resonator is capacitively coupled at a distance $l = 1.3$~mm, corresponding to a quarter-wavelength at frequency of approximately $f = 21$~ GHz.



An optical micrograph of the lumped-element resonator are shown in Fig.~\ref{fig:device}(c). The resonator is composed of an Al/AlO$_x$ Josephson junction (JJ) SQUID and interdigitated finger capacitors $C_\text{S}$. The lithographically defined coupling capacitors $C_\text{C}$ were designed to achieve a coupling quality factor $Q \sim$100, and corresponding coupling rate $\kappa\sim200$~MHz. In electromagnetic simulations, we found $\sim 5\%$ variation of $\kappa$ when varying the Josephson inductance $L_J$ from 90 to 200 pH. Crucially, we found that the junction capacitance $C_\text{J}$, often ignored in lower frequency applications, must be taken into account for higher-frequency designs. This is because $C_\text{J}$, which ranges between 290 to 130 fF for the $L_J$ range above, becomes comparable to $C_\text{S}\approx80$ fF as the targeted resonant frequency $f_r$ approaches the JJ plasma frequency, $f_p\approx 31$~GHz.

Additionally, the junction participation ratio $p = L/L_J$ is a crucial parameter for ensuring that this nonlinear resonator can operate reliably as a parametric amplifier for amplifying signals from qubits. Here $L$ corresponds to the effective inductance of the resonant mode, which includes the contribution of the non-negligible parasitic inductance ($L_S\approx140$ pH) of the shunt capacitors. As described in supplementary Sec. I, we estimate $p \approx 0.2$ from black-box quantization analysis of the simulated admittance~\cite{Manucharyan2007,Nigg2012,Frattini2018}. This gives a quality factor-participation ratio product $Q \cdot p \approx 20$, which ensures, according to previous  results,  that the device should reliably achieve gain $> 20$~dB~\cite{Manucharyan2007,narla2014,schackert2013practical}.

The key benefits of our WJPA design for high-frequency operation are three-fold: First, the waveguide launch simplifies packaging and reduces the losses present in circuit-board-based packages, which get worse at high frequency. Second, the wireless coupling scheme avoids impedance mismatches introduced by wire-bonds to the circuit board that also become problematic at high frequencies as they cause uncontrolled variation of $\kappa$. Third, the planar nature of the resonator and slotline transmission line eliminates manual alignment sensitivities that can cause variations in the coupling rate $\kappa$ from run to run. By integrating the resonator and slotline into a fully planar, lumped-element circuit, we decouple the design of the coupling interface from the surrounding three-dimensional package, enabling precise and repeatable control of $\kappa$. This planar approach also allows for incorporating on-chip lumped-element impedance matching networks in future iterations to broaden the amplifier bandwidth.

The WJPA chip is fabricated by first defining the slotline in tantalum (though other superconducting materials could also be used) on a sapphire wafer using photolithography and wet etching, followed by the fabrication of aluminum/aluminum-oxide Josephson junctions with the Dolan bridge technique. This $20 \times 20~\mu\text{m}^2$ SQUID loop (Fig.~\mbox{\ref{fig:device}}(d)) yields a total critical current $I_c$ of 3.7~$\mu$A, estimated from the room temperature resistance, corresponding to an effective Josephson inductance $L_J$ of approximately 90~pH at zero flux. During packaging, indium is applied around the chip edges to provide mechanical stability and suppress parasitic package modes.

\begin{figure}
    \centering
    \includegraphics[width=1\linewidth]{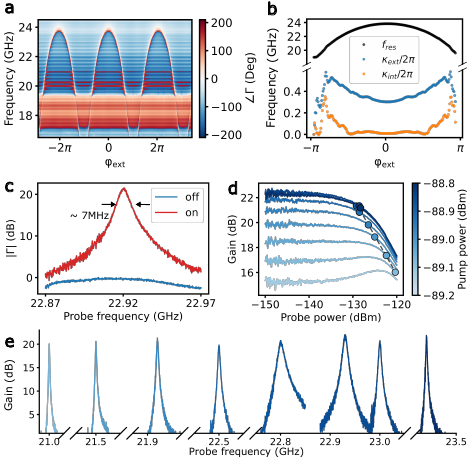}
    \caption{\textbf{WJPA linear response and gain.} 
    (a) Reflection coefficient phase as a function of external flux $\varphi_\text{ext}$.
    (b) Extracted resonant frequencies ($f_\text{res}$, black), external and internal coupling rates ($\kappa_\text{ext}$ in blue and $\kappa_\text{int}$ in orange) versus external flux.
    (c) Reflection coefficient when operated as a parametric amplifier with the pump applied (red), compared to the un-pumped response (blue).
    (d) Gain as a function of probe power for various pump powers, with extracted $P_\text{1dB}$ points (circles). A linear fit of $P_\text{1dB}$ versus pump power in dBm is shown by the gray dashed line whose slope is about 0.6.
    (e) Example gain profiles at various fluxes demonstrating frequency tunability from 21 GHz to 23.5 GHz.
    }
    \label{fig:gain}
\end{figure}

We first characterized the broadband linear response of the WJPA in a dilution refrigerator operated below 20~mK using a vector network analyzer (VNA). As shown in Fig.~\ref{fig:gain}(a), the resonance from the lumped-element nonlinear resonator was observed to be periodic with external flux, as expected. We identified a parasitic mode near 19.5~GHz with a linewidth of approximately 500~MHz whose position shifted across thermal cycles. We surmise that this mode originates from non-ideal sealing of the copper package. Additionally, we observed ripples in phase of about $\pm 25^{\circ}$ and in magnitude of about $\pm 1$~dB across the measured band (not shown), which we attribute to impedance mismatches at microwave interfaces between components in the setup. At each flux bias, the extracted resonant frequency and coupling rates are shown in Fig.~\ref{fig:gain}(b). Near zero flux bias, the resonator remains overcoupled with minimal variation in the coupling rate, consistent with simulations discussed in the supplementary Sec. I. As the flux bias is increased, the resonator approaches critical coupling near 19.5~GHz, likely due to the parasitic mode.

Next, we evaluate the amplification performance of the WJPA by measuring its reflection gain. The device was tested as a single-pump four-wave mixing (4WM) amplifier with a single microwave pump tone applied slightly below the bare resonance frequency to activate parametric gain. An example, shown in Fig.~\ref{fig:gain}(c), demonstrated  reflection gain over 20~dB with approximately 7~MHz bandwidth.


To characterize the amplifier's dynamic range, we measure gain compression as a function of input probe power for various fixed pump powers, as shown in Fig.~\ref{fig:gain}(d). The 1~dB compression points ($P_\text{1dB}$), marked with circles, is about -125 dBm at 20 dB of maximum gain. $P_\text{1dB}$ increases as the gain is decreased, and the extracted $P_\text{1dB}$ values decrease approximately linearly with pump power on a logarithmic scale, with a fitted slope of $\sim 0.6$. This trend aligns with theoretical analysis and prior experimental observations where gain saturation arises from Kerr-induced detuning and pump depletion~\cite{Eichler2014, Planat2019understand}. To further enhance the amplifier’s dynamic range and reduce pump-induced dephasing in qubit readout applications, a double pump configuration\mbox{ \cite{narla2014}} can be employed, which is compatible with the present design.

Finally, by tuning the magnetic flux through the SQUID loop, the resonant frequency of the amplifier is continuously adjusted. As demonstrated in Fig.~\ref{fig:gain}(e), the WJPA supports parametric gain exceeding 20~dB across a frequency range from 21~GHz to 23.5~GHz, with typical bandwidth of 3~MHz. This result highlights its suitability for aligning over a broad frequency range with a readout resonator coupled to a qubit.


Characterization of the added noise of a Josephson amplifier is crucial for its application in qubit readout. In this work, we implemented two methods to characterize the noise performance of the WJPA and cross checked the results. The experiment wiring configurations for both methods can be found in supplementary Sec.~II. In both methods, we quote the noise assuming phase-preserving mode of amplification, with the pump frequency detuned from the measurement frequency, for which the ideal amplifier would have added noise $N_\text{add} = 1/2$ photon\cite{caves1982quantum}.

\begin{figure}
    \centering
    \includegraphics[width=1\linewidth]{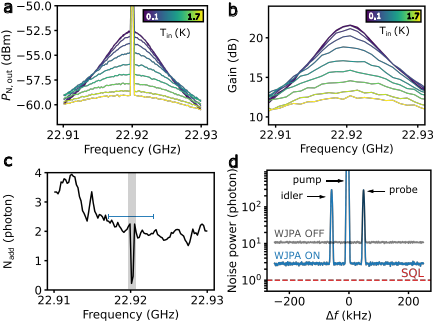}
    \caption{\textbf{WJPA noise performance.} 
   (a) Measured output noise power $P_{N,\mathrm{out}}$ as a function of spectrum analyzer (SA) frequency for various temperatures of the VTS (colorbar). The peak near the JPA center frequency corresponds to the pump tone seen on the spectrum analyzer.
(b) WJPA gain measured by the VNA at corresponding temperatures of the VTS in part (a).
    (c) Extracted JPA-added noise $N_{\mathrm{add}}$ at the reference plane of the VTS. A blue bar marks the amplifier bandwidth of $7$ MHz.  A dip near the JPA center frequency (gray area) corresponds to the pump tone.
    (d) Power spectrum referred to the input of the WJPA. The gray trace corresponds to the unpumped case (WJPA OFF). The blue trace shows the spectrum with the WJPA pump turned on. The central peak is the pump tone, and the adjacent upper side peak is the probe tone injected by a separate microwave generator while the lower side peak of the blue trace is the idler tone. The red dashed line indicates the standard quantum limit (SQL) at 22~GHz. Both traces were acquired with a spectrum analyzer resolution bandwidth of 4.7~kHz.
    }
    \label{fig:noise}
\end{figure}

We start by using the Y-factor technique with a variable temperature source (VTS)~\cite{Malnou2024low}, consisting of a 50~$\Omega$ matched microwave attenuator thermally anchored to a copper bracket with a heater and ruthenium oxide thermometer, and thermally isolated from the mixing chamber stage of the dilution refrigerator. By varying the VTS temperature, thermal noise is generated corresponding to a known photon occupation number $N_\text{in}(T_\text{in}) = {1}/({e^{hf/k_B T_\text{in}} - 1}) + 1/2$, where $h$, $k_B$ are Planck and Boltzmann constants, $f$ is the frequency and $T_\text{in}$ is the attenuator temperature. Fig.~\ref{fig:noise}(a) shows the measured output noise power $P_{N,\mathrm{out}}$ versus frequency for various $T_\text{in}$. Unlike the expectation for an ideal linear amplifier where $P_{N,\mathrm{out}}$ would increase with $T_\text{in}$, we observe a reduction in output noise with increasing temperatures. As discussed in Ref.~\onlinecite{Malnou2024low}, this behavior arises because of gain compression, from the broadband Johnson noise at elevated $T_\text{in}$ saturating the amplifier and reducing its gain.

This gain compression is visualized in Fig.~\ref{fig:noise}(b), where we plot the gain versus frequency measured by the VNA for each $T_\text{in}$. The gain decreases with increasing $T_\text{in}$, dropping from 21.6~dB at 0.1~K to 12.9 dB at 1.75~K. To correctly extract the added noise, we account for this gain reduction by scaling the noise spectrum with the measured gain following the procedure discussed in Ref.~\onlinecite{Malnou2024low}, and further described in supplementary Sec. III. The resulting WJPA-added noise $N_\text{add}$ as a function of frequency is shown in Fig.~\ref{fig:noise}(c). It shows that across the amplifier bandwidth, $N_\text{add}$ is approximately 2 photon. It is worth noting that this value likely overestimates the intrinsic JPA performance due to the lossy stainless steel cable ($\sim$2.1~dB at 22 GHz, see wiring details in Supplementary Fig. S2) connecting the VTS to the WJPA. Because the reference plane for the Y-factor analysis is defined at the VTS, any attenuation from the intervening cable effectively appear as additional added noise. The finite temperature gradient across the cable length also makes the reference plane ambiguous. Considering these effects, as well as other approximations discussed in Supplementary Sec. III, we quote $N_\text{add} = 2$ photons as a conservative result of the WJPA-added noise.

Next, we implemented a qubit-based noise calibration approach using a transmon-cavity system~\cite{macklin2015near,Bultink2018general} with a readout cavity frequency of $21.7$~GHz. Details of this method are described in supplementary Sec.~IV. Fig.~\ref{fig:noise}(d) shows the calibrated noise spectrum referenced to the input of the WJPA, acquired at 22~GHz. When the WJPA is turned off, the noise power is dominated by the HEMT noise, measured to be approximately 10.3 photons, or system noise temperature $T_\text{sys} =10.9$~K at 22~GHz. With the WJPA turned on, the total system added noise drops to $N_\text{sys} = 2.3$ photon, or system noise temperature $T_\text{sys} = 2.4$~K. Note that $N_\text{sys}$ combines the noise contributions of losses from cables and components between the transmon-cavity and WJPA, the WJPA-added noise $N_\text{add}$, and the residual added-noise of the rest of the amplification chain after the WJPA. We can also express these results in terms of a system quantum measurement efficiency\mbox{\cite{Mallet2011quantum,white2023}} $\eta =1/(1 + 2N_\text{sys} )\approx 18\%$ with the WJPA turned on, compared to 4.6\% with the WJPA turned off.

To summarize, both noise calibration methods indicate the system added noise with the WJPA operating of approximately 2 photon. This value is likely limited by losses in coaxial cables and circulators between the VTS/qubit measurement plane and the WJPA port, which could be improved with high-frequency microwave system engineering in the future. Overall, this noise performance confirms the WJPA’s capability to operate near the quantum limit and substantially improve fidelity of measurement in high-frequency readout applications.

In summary, we have demonstrated a wireless, waveguide-coupled Josephson parametric amplifier (WJPA) that features a lumped-element resonator embedded with a JJ SQUID and coupled to the input/output port via a lithographically defined waveguide-to-slotline coupling structure. Operated in four-wave mixing mode with a single pump tone, the WJPA achieves over 20~dB of gain across a tunable range from 21~GHz to 23.5~GHz, with typical 3~MHz bandwidth. Using both Y-factor measurements and a qubit-based noise calibration, we extract a system added noise of approximately two photons.

Our work opens up a few natural next steps. The measured P$_\text{1dB}=-125$~dBm at the operating frequency of 22.9~GHz can be alternatively expressed as 21 photons/$\mu$s. This number should be compared to typical drive power values used for fast, high-fidelity qubit readout $\sim 100$~photons/$\mu$s\mbox{\cite{Heinsoo2018,Hazra2025,Spring2025}}. If as expected, similar drive powers are also needed for high frequency qubit readout, then amplifier compression could pose a limit to readout fidelity. Thus a natural next step to our work would be to characterize the fidelity and repeatability of high-frequency dispersive qubit readout with such an amplifier. We also note that strategies to increase amplifier compression power by arraying junctions and increasing their critical current\mbox{\cite{Eichler2014,Frattini2018,Kaufman2025Simple}}, as well as increasing bandwidth through impedance engineering\mbox{\cite{Naaman2022}}, should be adaptable to higher operating frequency as well, as long as attention is paid to the limits imposed by the junction plasma frequency.

Furthermore, while this work realized a WJPA with Al/AlO$_x$ junctions, the design is readily implementable with Josephson elements realized with superconductors such as niobium that can operate at high temperature $\sim 1$~K\mbox{\cite{anferov2024millimeter}}, as well as kinetic inductance materials like niobium nitride\mbox{\cite{Anferov2020,xu2023magnetic}}, NbTiN\mbox{\cite{Parker2022deg,hung2025broadband}}, and granular aluminum\mbox{\cite{Maleeva2018GrAl,Zapata2024granular}}. In particular, we note that self-resonance frequencies in excess of 70~GHz have been measured with granular aluminum\mbox{\cite{Maleeva2018GrAl}}. Characterizing such self-resonance frequencies for other Josephson elements and kinetic inductance materials will be especially helpful in understanding the limits on designing amplifiers for higher operation frequencies.

\section*{Supplementary Material}
See the supplementary material for details on the design and microwave simulations of the WJPA, as well as the noise measurement setups and methods, including both the Y-factor technique and the qubit-based approach.

\section*{Acknowledgment}
This work was supported by the Defense Advanced Research Projects Agency (Grant No. HR00112420343), the Army Research Office (Grant No. W911NF-23-1-0251 and W911NF-23-1-0096) and the Air Force Office of Scientific Research (Grant No. FA9550-22-1-0203). HMC was supported in part by an appointment to the Department of Defense (DOD) Research Participation Program
administered by the Oak Ridge Institute for Science and Education (ORISE) through an interagency agreement between the U.S.
Department of Energy (DOE) and the DOD. ORISE is managed by Oak Ridge Associated Universities (ORAU) under DOE contract number DE-SC0014664. Sample fabrication was performed in the University of Texas at Austin Microelectronics Research Center, a member of the National Nanotechnology Coordinated Infrastructure (NNCI), which is supported by the National Science Foundation (Grant No. ECCS-2025227). All opinions expressed in this paper are the author's and do not necessarily reflect the policies and views of DOD, DOE, or ORAU/ORISE.
\section*{DATA AVAILABILITY}
The data that support the findings of this study are openly available on Zenodo at doi.org/10.5281/zenodo.15657612.
\bibliography{manuscript}

\end{document}


\preprint{AIP/123-QED}

\title[Supplementary for ``Wireless Josephson parametric amplifier above 20 GHz'']{Supplementary for ``Wireless Josephson parametric amplifier above 20 GHz''}
\author{Z. Hao}
\author{J. Cochran}
\author{Y.-C. Chang}
\author{H. M. Cole}
\author{S. Shankar}
\affiliation{Chandra Department of Electrical and Computer Engineering, University of Texas at Austin, 2501 Speedway, Austin, TX 78712, USA}
 \email{shyam.shankar@utexas.edu}

\date{\today}

\renewcommand{\figurename}{Fig.}
\renewcommand{\thefigure}{S\arabic{figure}}
\renewcommand{\thetable}{S\arabic{table}}
\renewcommand{\theequation}{S\arabic{equation}}
\maketitle
\section{\label{sec:bbq}Numerical simulation of the WJPA}

\begin{table}
    \centering
    \begin{ruledtabular}
    \begin{tabular}{ccc}
        Symbol & Value & Definition \\
        $L_J$ & 120 pH& Junction inductance \\
        $C_J$ & 220 fF& Junction capacitance \\
        $C_S$& 80 fF & Shunt capacitance \\
        $C_C$& 30 fF& Coupling capacitance \\
        $L_S$ & 140 pH& Parasitic inductance in series with $C_\text{S}$ \\
        $W_a$ & 4.32 mm & Smaller rectangular waveguide dimension \\
        $S$ & 200 $\mu$m & Slotline gap width \\
        $A$ & 4.5 mm & Taper transition length \\
        $l$ & 1.3 mm & Resonator distance from the short \\
        $g_g$ & 3 $\mu$m & Shunt capacitor finger gap \\
        $l_g$ & 60 $\mu$m & Shunt capacitor finger length \\
        $w_g$ & 10 $\mu$m & Shunt capacitor finger width \\
        $N_g$ & 7 & Number of shunt capacitor fingers \\
        $g_c$ & 6 $\mu$m & Coupling capacitor finger gap \\
        $l_c$ & 20 $\mu$m & Coupling capacitor finger length \\
        $w_c$ & 21 $\mu$m & Coupling capacitor finger width \\
        $N_c$ & 4 & Number of coupling capacitor fingers \\
    \end{tabular}
    \end{ruledtabular}
    \caption{WJPA circuit design parameters and taper design parameters.}
    \label{tab:t1}
\end{table}

\begin{figure}
    \centering
    \includegraphics[width=1\linewidth]{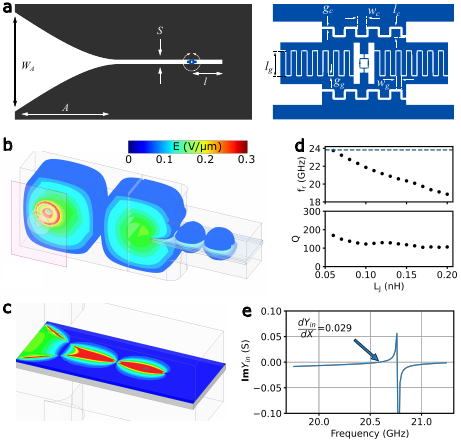}
    \caption{\textbf{Electromagnetic simulation of the WJPA.} 
    (a) Top view of the WJPA chip geometry, with the detailed view of the lumped element resonator region (circled with dotted line) on the right. Image to scale.
    (b) Simulated electric field profile of the WJPA at approximately 20~GHz showing coupling from the waveguide to the on-chip slotline. 
    (c) On-chip field distribution highlighting resonator placement where the E-field is near the maximum. (b) and (c) share the same color bar indicating the E-field strength with port excitation of 1 watt.
    (d) Extracted resonant frequency $f_\text{res}$ and quality factor $Q$ as functions of the Josephson inductance $L_J$. Measured $f_\text{res}=23.8 $ GHz of the device at zero flux is marked with a blue dotted line.
    (e) Black-box quantization simulation to extract participation ratio $p$ with the Josephson inductance $L_J$ fixed at 120~pH.}
    \label{fig:wjpa-sim}
\end{figure}

We first designed and simulated the lumped element circuit schematic model of the WJPA, shown in Fig.~1, in AWR Microwave Office to target a resonance frequency of about $22$~GHz with quality factor of 100 for a junction with $L_J\sim120$~pH. Assuming a junction critical current density of $70$~A/cm$^2$ and specific capacitance of 55~fF/$\mu$m$^2$, we find that the junction capacitance $C_J\sim 220$~fF is not negligible, thus it is included in the simulation. Under these constraints, the design yielded target values of  $C_S\sim 100$~fF, $C_C\sim36$~fF. 

Subsequently, full-wave electromagnetic (EM) simulations of the WJPA resonator were performed in Ansys HFSS. The interdigitated capacitors (IDCs) were initially designed using empirical formulas and then iteratively refined until the simulated resonance frequency and linewidth matched the targets. The finalized IDC geometries and parameter symbols are illustrated in Fig.~\mbox{\ref{fig:wjpa-sim}}(a), and their dimensions are summarized in Table~\mbox{\ref{tab:t1}}. Finally, the shunt capacitor $C_S$ and its parasitic inductance $L_S$, as well as the coupling capacitor $C_C$ were individually simulated in AWR Microwave Office, and results are reported in Table~\mbox{\ref{tab:t1}}.

The three-dimensional simulation included the WR42-to-coaxial adapter, copper waveguide cavity, and the sapphire chip with the on-chip lumped-element resonator placed on the slotline [Fig.~\ref{fig:wjpa-sim}(b,c)].  The Josephson junction is modeled as a linear inductor. Using eigenmode analysis, we extract the resonant frequency $f_\text{res}$ and corresponding quality factor $Q$ as functions of the junction inductance $L_J$. As shown in Fig.~\ref{fig:wjpa-sim}(d), $f_\text{res}$ decreases with increasing $L_J$ while $Q$ remains relatively stable across the sweep. 

The device is estimated to have $L_J = 90$ pH at zero flux from the room-temperature resistance measurement. The simulated resonance frequency $f_{\mathrm{res}} = 22.3~\mathrm{GHz}$ at this inductance is about 6\% lower than the measured zero-flux value ($23.8~\mathrm{GHz}$). This offset likely arises from fabrication variations in the shunt capacitance and contact resistance in the two-probe resistance measurement, which can slightly overestimate $L_J$.



To evaluate the participation ratio $p$ of the junction using black-box quantization (BBQ)~\mbox{\cite{Manucharyan2007}}, we extract the simulated input admittance $Y_\text{in}$ at the port [Fig.~\mbox{\ref{fig:wjpa-sim}}(e)]. The slope $\partial \mathrm{Im}(Y_\text{in})/\partial \omega$ at resonance gives the mode capacitance $C \approx 2.3$~pF and the corresponding linear resonant mode inductance $L \approx 26$~pH. From these we obtain a participation ratio $p = L/L_J \approx 22\%$ for $L_J = 120$~pH.

\section{\label{sec:noise setup}Noise measurement setups}
\begin{figure}
    \centering
    \includegraphics[width=1\linewidth]{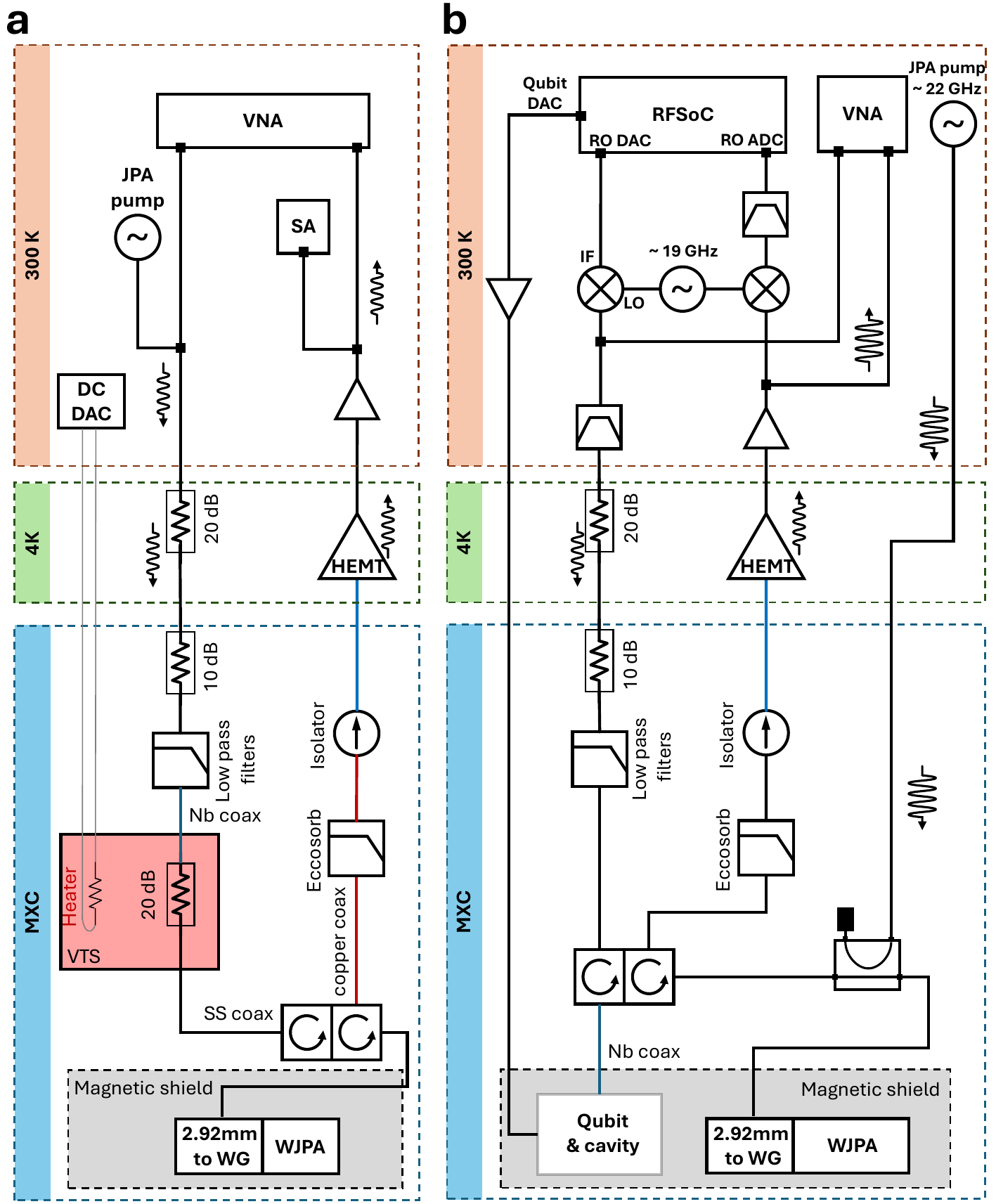}
    \caption{\textbf{Wiring diagrams for WJPA noise measurements.}
Experiment setup for the (a) Y-factor, and (b) qubit-based calibration of the added noise of the WJPA. 
}
\label{fig:wiring}
\end{figure}
Two experiment setups are used in the noise measurement. First, the Y-factor measurement setup is depicted in Fig.~\ref{fig:wiring}(a). A vector network analyzer (VNA, Keysight E8364A) is used to perform linear reflection coefficient and gain measurements. The WJPA pump is provided by a microwave generator (SignalCore SC5521A), then combined at room temperature with the VNA probe tone using a power combiner (Minicircuits ZC2PD-18265-S+). The input filters consist of a home-made eccosorb filter and a commercial suspended-substrate low-pass filter (Minicircuits ZLSS-K24G+). A home-made variable temperature stage (VTS), shown in the red box, consists of a 50~$\Omega$ matched load (20~dB attenuator, XMA 4882-6240-20-CRYO) mounted on a thermally isolated copper support, which includes an cartridge heater and a calibrated thermometer. It is weakly thermally anchored to the mixing chamber plate, allowing its temperature to be varied independently from the base temperature of the dilution refrigerator. In our setup, the maximum temperature that the VTS can stabilize is 1.75 K.  A stainless steel cable is used to thermally isolate the VTS and the circulator (Quinstar QCY-G1802652AM-R-1). The WJPA package with the waveguide-to-coax adapter (Fairview Microwave FMWCA1019) is installed inside a cryogenic $\mu$-metal shield. The output line consists of a home-made eccosorb filter, isolator (Quinstar QCI-G1802652AM-R-1), 15 $\sim$ 29~GHz HEMT amplifier (Low Noise Factory LNF-LNC15-29B) and two room temperature amplifiers (Eravant SBL-1832734025-KFKF-S1). A spectrum analyzer (Keysight N9000B) measures the output noise while varying the VTS temperature.

The qubit-based noise calibration setup shown in Fig.~\ref{fig:wiring}(b), shares most components with the Y-factor measurement setup. Specific differences are as follows: (1) Control signals are generated by up/down-converting waveforms from a high-speed FPGA (Xilinx ZCU216 RFSoC with QICK firmware\onlinecite{Ding2024qick}) to K-band via mixers (Minicircuits ZMDB-44H-K). (2) A directional coupler (Krytar 262220) is used to deliver the pump tone for the WJPA.

\section{\label{sec:noise y-factor}Noise measurement of the WJPA using the Y-factor method}
To clarify the multiple symbols used for noise terms discussed below and in the main text, we briefly introduce them first. $N_\text{in}$ denotes the noise incident on the amplifier input, determined solely by the components preceding the amplifier (e.g., attenuators and cables). $N_\text{sys}$ represents the system-added noise referred to the input of the entire amplification chain, corresponding to the noise added by an equivalent single amplifier that captures the combined effect of all amplifier stages. $N_\text{add}$, in contrast, denotes the added noise contributed only by the first-stage amplifier—the WJPA in our case. Because the WJPA provides high gain ($\sim 20$~dB), it suppresses most of the subsequent HEMT noise, making $N_\text{add}$ slightly smaller than but close to $N_\text{sys}$.

In the main text, the $N_\text{add}$ values are reported from the VTS-based noise calibration method (Fig.~3(c)), where the formulas below naturally yield the first-stage added noise. In contrast, $N_\text{sys}$ is quoted from the qubit-based calibration method (Fig.~3(d)), which measures the total system noise without resolving individual amplifier contributions.

The output noise with the WJPA ON decreases with increasing input noise due to  gain saturation. Moreover the observed output noise of a narrow-band parametric amplifier has contributions from the noise from the VTS incident on the device at both signal and idler frequency. This behavior requires special treatment to accurately characterize the added noise of the WJPA, following the procedure in Ref.~\onlinecite{Malnou2024low}. We can write the total noise output by the amplifier chain as
\begin{align}
N_\text{out} &=G_\text{sys, rest} \left[G_\text{WJPA}(N_\text{in} + N_\text{in}^i + N_\text{add, ex})+ N_\text{sys, rest}\right].
\end{align}
$G_\text{sys, rest}$ and $N_\text{sys, rest}$ are the gain and noise of the amplifier chain following the WJPA. $G_\text{WJPA}$ is the WJPA gain. $N_\text{in}$ and $N_\text{in}^i$ are the input noise incident on the WJPA at the signal and idler frequencies. Following Ref.~\onlinecite{Malnou2024low}, we assume  $  N_\text{in}$, $ N_\text{in}^i$ are given by the quantum Johnson noise formula $N_\text{in}^\text{Johnson}= (1/2)\text{coth}\left(hf/2 k_B T_\text{VTS}\right)$, where $h$ is the Planck's constant, $k_B$ is the Boltzman constant, $f$ is the measurement frequency and $T_\text{VTS}$ is the VTS temperature. $N_\text{add, ex}$ refers to excess JPA added noise referred to the WJPA plane, such as due to internal loss in the device or transmission losses in the package.

Assuming $  N_\text{in} = N_\text{in}^i $ and that the excess added noise of the WJPA remains constant during the VTS temperature sweep, we renormalize the output noise by dividing by the WJPA gain:
\begin{align}
\frac{N_\text{out}}{G_\text{WJPA}} &= 2 G_\text{sys, rest}\left(N_\text{in} + \frac{N_\text{add, ex}}{2} +\frac{N_\text{sys, rest}}{2G_\text{WJPA}}\right).
\end{align}
We measured $N_\text{sys, rest}\approx 20$ in a separate Y-factor measurement on the amplifier chain with the WJPA OFF, dominated by the added noise of the HEMT amplifier. It is worth noting that this value represents an upper bound on the intrinsic HEMT noise, since the finite loss of the stainless-steel cable (approximately 2.1 dB) effectively increases the noise when referred to the input plane\mbox{\cite{Malnou2024low}}. Accounting for this loss, we get the HEMT added noise of 12.1 photons, close to the 10.3 photons result from the qubit-based method  where the lossy stainless steel cable was replaced. Neglecting the last term for the moment, we can extract $N_\text{add,ex}$ from a linear regression fit of $N_\text{out}/G_\text{WJPA}$ versus $N_\text{in}$. The procedure gives the extracted JPA-added noise as $N_\text{add} = N_\text{add, ex} + 1/2 $ where $1/2$ corresponds to the minimum quantum Johnson noise incident on the JPA at the idler frequency. $N_\text{add}$ is shown as a function of frequency, as shown in the main text Fig.~3(c). 

Because the stainless-steel cable introduces attenuation before the WJPA, the $N_\text{add} = 2$ photons should be regarded as an upper bound on the WJPA’s intrinsic added noise. Correcting for the cable loss yields an estimated WJPA-added noise of approximately 1.1 photons. Nevertheless, we report the more conservative value of 2 photons in the main text.

For $G_\text{WJPA} \approx 20~\mathrm{dB}$, the term $\tfrac{N_\text{sys,rest}}{2 G_\text{WJPA}} \approx 0.1$ corresponds to about 10\% of the $\tfrac{N_\text{add,ex}}{2}$ contribution, resulting in a slight overestimation of $N_\text{add}$. It is difficult to fully eliminate this correction because $G_\text{WJPA}$ varies with frequency. Consequently, the accuracy of estimating $N_\text{add}$ by this method decreases at elevated VTS temperatures and for frequencies outside the WJPA bandwidth, where the gain drops to approximately $12~\mathrm{dB}$.




\section{\label{sec:cQED}Noise measurement of the WJPA with cQED device}

The noise measurement with a qubit is based on calibrating the power of a reference coherent drive at the port of the readout resonator coupled to the qubit~\cite{macklin2015near}. The calibrated drive power is then used along with the measured power spectrum on the spectrum analyzer to estimate the system gain and noise.

\begin{figure}
    \centering
    \includegraphics[width=1\linewidth]{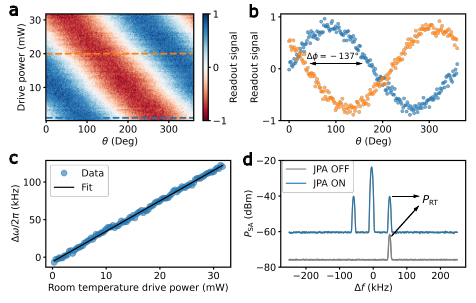}
    \caption{
\textbf{Drive induced Stark shift.}     (a) Readout signal as a function of second qubit pulse phase $\theta$ and drive power, showing periodic interference fringes whose phase shifts with the applied room temperature drive power. 
    (b) Extracted signal amplitudes as a function of $\theta$ at two drive powers corresponding to dashed lines marked on (a).
    (c) Data and linear fit of Stark shift $|\Delta \omega/2\pi|$ versus drive power at room temperature.
    (d) Power spectrum measured on a room temperature spectrum analyzer (SA) of the output power level with the WJPA pump turned on (blue trace) and off (gray trace). Arrows mark the power $ P_\text{RT}$ at the drive frequency $\omega_d$. X axis is the frequency offset from the pump center at 22 GHz.
}
    \label{fig:qubit-weak}
\end{figure}
We used a cQED device that consists of a transmon qubit dispersively coupled to the $\text{TE}_{101}$ mode ($\omega_r/2\pi = 21.765\,\mathrm{GHz}$) of a rectangular cavity. The qubit had frequency $\omega_q/2\pi=6.7$~GHz,  dispersive coupling to the readout mode $\chi/2\pi=9.6$~MHz and coherence times $T_1\approx20$~$\mu$s and $T_{2R}\approx10$~$\mu$s. A Ramsey-experiment is used to measure the Stark-shift of the qubit frequency in the presence of a weak continuous drive. The Ramsey experiment consisted of two $\pi/2$ pulses, $R_X(\pi/2)$ and a $R_\theta(\pi/2)$ separated by a fixed delay $\tau = 5$~$\mu$s, followed by a qubit measurement. The phase $\theta$ of the second pulse was swept from 0 to $2\pi$. Additionally, a continuous drive with variable power was applied at $\omega_d/2\pi = 22\,\mathrm{GHz}$, detuned from the cavity resonance. Under these conditions, we expect a pronounced Stark shift and negligible dephasing of the qubit. This Stark shift appears as a phase shift of the Ramsey interference pattern [Fig.~\ref{fig:qubit-weak}(a,b)], from which the Stark shift $|\Delta \omega|$ is extracted [Fig.~\ref{fig:qubit-weak}(c)] using the relation $\Delta\phi = \tau \Delta \omega$, where $\Delta\phi$ is the observed phase shift in radians. 

The measured Stark shift for any room temperature drive power, can be linked to the power at the plane of the readout cavity as follows. From Ref.~\onlinecite{gambetta2008}, we have
\begin{align}
    \Delta \omega &= \chi \, \mathrm{Re} \left\{ \alpha_e^* \alpha_g \right\},
 \label{eq:Gamma_m}
\end{align}
where $\chi$ is the dispersive shift of the qubit-cavity system and the complex amplitudes $\alpha_g$ and $\alpha_e$ represent the steady-state intra-cavity amplitude when the qubit is in the ground or excited state, respectively. $\alpha_g$, $\alpha_e$ can be expressed as
\begin{align}
    \alpha_g &= \frac{-j \epsilon_d}{\kappa/2 + j (\Delta_r )}, \\
    \alpha_e &= \frac{-j \epsilon_d}{\kappa/2 + j (\Delta_r - \chi)},
\end{align}
where $\Delta_r = \omega_r - \omega_d$ is the detuning between the cavity and the drive frequency, $\kappa$ is the cavity linewidth and $\epsilon_d$ is the drive amplitude incident on the readout cavity. $\epsilon_d$ is related to the power entering/exiting the cavity by $P_\text{in,out} \approx \frac{\hbar\omega_d}{\kappa}|\epsilon_d|^2$ following the derivation in Ref.~\onlinecite{clerk2010}, when the drive is far off-resonance ($\Delta_r\gg\chi,\kappa$), and assuming no internal loss in the cavity. With these relations, we can use the qubit effectively as a calibrated power meter by linking the measured Stark-shift for any drive power to the absolute power $P_\text{in,out}$ at the readout cavity. 

Next, we measure the corresponding output power at the drive frequency $\omega_d$ on the room temperature spectrum analyzer ($ P_\text{RT}$), shown in Fig.~\ref{fig:qubit-weak}(d), to estimate the system gain $G_\text{sys} = P_\text{RT}/P_\text{in,out}$, resulting in $G_\text{sys} = 96.8$ dB when JPA is turned on and $75.5$ dB when off, at $\omega_d/2\pi = 22$~GHz. Finally, the measured noise power spectrum $P_\text{SA}\left(\Delta f\right)$ is expressed in photon number units, shown in the main text Fig.~3(d) by the relation
\begin{align}
   N\left(\Delta f\right) = \frac{P_\text{SA}\left(\Delta f\right)}{G_\text{sys}\hbar\omega_dr_\text{BW}},
\end{align}
where $r_\text{BW}=4.7$~kHz was the resolution bandwidth on the analyzer.



\bibliography{supp_reference}